\documentclass[reprint,prd,aps]{revtex4-1}
\usepackage{amsmath}
\usepackage{amssymb}
\usepackage[]{graphicx}
\usepackage[caption=false]{subfig}
\usepackage{hyperref}

\begin{document}

\title{{Graviton production from the CMB in large scale magnetic fields}}
\author{Damian Ejlli}
\email{ejlli@fe.infn.it}

\affiliation{Dipartimento di Fisica e Scienze della Terra, Universit\'{a} degli Studi di Ferrara, \\
Polo Scientifico e Tecnologico-Edificio C, Via Saragat 1, 44122 Ferrara, Italy}
\affiliation{Astroparticule et Cosmologie (APC),
Universit\'{e} de Paris Diderot-Paris 7\\
10, Rue Alice Domon et L\'{e}onie Duquet,
75205 Paris Cedex 13
France}

\date{\today}
\begin{abstract}

Conversion of the CMB photons into gravitational waves at the post recombination epoch is considered. We calculate the probability of transformation of the CMB photons into gravitons in the presence of a large scale magnetic field. Based on the present day limits on the strength of the large scale magnetic field, we show that the probability of the produced gravitons as a result of photon to graviton conversion is reasonable and such a mechanism would produce an isotropic background of gravitational waves in the same frequency range of the CMB photons. The mechanism proposed would be another good opportunity to study the high frequency part of the spectrum of gravitational waves.
\end{abstract}

\maketitle

\section{Introduction} 
\label{sec:Introduction}

One of the most important predictions of the standard cosmology is without any doubt the cosmic microwave background (CMB) radiation which gives us important information about the state of the Universe at the time of the last scattering. The CMB has a perfect blackbody spectrum and presents very small temperature anisotropy of the order of $\delta T/T\sim 10^{-5}$. The temperature anisotropy can be divided into two groups: primary anisotropy, due to effects which occur at the last scattering surface and before; and secondary anisotropy, due to effects such as interactions of the background radiation with hot gas or gravitational potentials, which occur between the last scattering surface and the observer.

In connection with the secondary anisotropy, in this work we study the interaction of the CMB photons with large scale magnetic field at the post recombination epoch. The study of this kind of interaction is twofold; first the interaction of the CMB with the large scale magnetic field can in principle produce gravitons through the mechanism of photon to graviton conversion in the magnetic field \cite{gertsen} and, second, the produced gravitons might in principle cause small temperature anisotropy. 

The inverse process, namely, the mechanism of graviton to photon conversion in a magnetic field, was first considered in \cite{g-to-gamma}. Conversion of high energy gravitons into photons at the post recombination epoch was considered in \cite{Dolgov:2012be, Dolgov:2013pwa} where a substantial background of x-ray photons might be created at the post recombination epoch. The origin of such high energy gravitons has been studied in \cite{hi-f-gw} and is thought to be emitted by a population of primordial black holes through the mechanism of Hawking evaporation.

Conversion of the CMB photons into gravitons at the post recombination epoch was first studied in  \cite{Chen:1994ch} in the wave function approximation and in \cite{Magueijo:1993ni} at a classical level. In \cite{Chen:1994ch, Magueijo:1993ni} it was argued that for the present day values of the magnetic field such a mechanism would produce the observable CMB anisotropy. Later in \cite{Cillis:1996qy} the plasma effects were included in the oscillation probability and it was shown that the photon to graviton oscillation probability is negligible in order to produce the CMB temperature anisotropy.

In this work we go beyond the wave function approximation and consider coherence breaking of the photon to graviton oscillation in plasma. The frequency range of the formed gravitons is the same as the frequency range of the CMB photons and makes it hard to detect by the next generation of gravitational wave (GW) detectors. However, we use such an opportunity and put it into a detection prospective for the future high frequency GW detectors.

 \section{Photon-graviton mixing}
 \label{sec:Photon-graviton mixing}
 
 
In the limit where the photon wavelength $\lambda_p$ is smaller than the coherence length of the magnetic field $\lambda_B$, $\lambda_p\ll \lambda_B$, the equations of motion of the photon-graviton system are very well described by the WKB approximation and therefore are reduced to the following matrix equation \cite{Dolgov:2012be, Raffelt:1987im}:
\begin{equation}
\left[(\omega+i\partial_{\mathbf{x}})\mathbf I+
\begin{bmatrix}
  \omega(n-1)_{\lambda} & B_{T}/m_{Pl} \\
  B_{T}/m_{Pl}  & 0  \\
   \end{bmatrix}
\right] 
\begin{bmatrix}
  A_\lambda({\mathbf{x}}) \\
  h_\lambda({\mathbf{x}}) 
 \end{bmatrix}
=0\,,
\label{matrix}
\end{equation}
where $\mathbf I$ is the unit matrix, $\mathbf x$ is the direction of the graviton/photon propagation,
$n$ is the total refraction index of the medium, $\omega$ is the graviton energy, $B_T$ is the strength of the transverse external magnetic field $\mathbf B_e$ and $h_\lambda, A_\lambda$ are  respectively the graviton and photon polarization states with $\lambda$ being the polarization index (helicity) of the graviton and the photon states. In the case of photons $\lambda=+$ indicates a polarization state perpendicular to the external magnetic field and $\lambda=\times$ indicates a state with polarization parallel to the external field. 
 
 The refraction index of the photons is given by the sum of three terms,
 \begin{equation}
n=n_\textrm{pl}+n_\textrm{QED}+n_\textrm{CM},
\end{equation}
where $n_\textrm{pl}$ is the index of refraction due to a plasma medium; $n_\textrm{QED}$ is the index of refraction due to vacuum polarization; and $n_\textrm{CM}$ is the index of refraction due to the Cotton-Mouton effect. The refraction index due to plasma effects is given by $n_\textrm{pl}=-\omega_\textrm{pl}^2/2\omega^2$ with $\omega_\textrm{pl}^2=4\pi\alpha n_e/m_e$ where $n_e$, here, is the plasma density and $m_e$ is the electron mass. The Cotton-Mouton effect gives rise of two different indices of refraction for each photon polarization state where $n_\textrm{CM}^+-n_\textrm{CM}^\times=C\lambda_p B_e^2$ with $C$ being the Cotton-Mouton constant. In this work we neglect the contribution of Cotton-Mouton effect to 
the refraction index due to difficulties in determining the Cotton-Mouton constant $C$ at the post recombination epoch.The index of refraction due to vacuum polarization in the case of $\omega\ll (2m_e/3)(B_c/B)$ reads \cite{Tsai:1974fa}
\begin{equation}\label{QED-index}
n_{\times, +} =1+\frac{\alpha}{4\pi}\left(\frac{B_T}{B_c}\right)^2\left[\left(\frac{14}{45}\right)_{\times}, \left(\frac{8}{45}\right)_{+}\right].
\end{equation}
where $B_T=B_e\sin\Theta$ with $\Theta$ being the angle between $\mathbf B_e$ and $\mathbf x$; and $B_c=m_e^2/e=4.41\times 10^{13}$ G.

Since photons do oscillate and scatter during their propagation in the medium, in this work we work in the density matrix formalism. The density operator satisfies the Liouville-von Neumann equation
\begin{equation}
\frac{d \hat\rho}{d t }= -i[\hat H, \hat\rho],
\label{densityoperator}
\end{equation}
where $\hat H=\hat M-i\hat\Gamma$ is the total Hamiltonian of the system which is in general not Hermitian. The operators $\hat M$ and $\hat\Gamma$ take into account respectively, refractive and damping processes of the system with the medium. Their matrix representations are given, respectively,
\begin{equation}
M_\lambda=
\begin{bmatrix}
m_\lambda & m_{g\gamma}\\
m_{g\gamma} & 0
\end{bmatrix},
\quad 
\Gamma=
\begin{bmatrix}
\Gamma_\gamma & 0\\
0 & 0
\end{bmatrix}
\end{equation} 
where $M_\lambda=\langle\Psi_\lambda|\hat M|\Psi_\lambda\rangle$, $\Gamma=\langle\Psi_\lambda|\hat \Gamma|\Psi_\lambda\rangle$ with $\Psi_\lambda^T=[A_\lambda, h_\lambda]$ and $T$ means the transpose of a given matrix. The matrix elements of $M_\lambda$ are respectively $m_\lambda=\omega(n-1)_\lambda$ and $m_{g\gamma}=B_T/m_\textrm{Pl}$. In the matrix $\Gamma$ the term $\Gamma_\gamma$ indicates the interaction rate of photons with the medium. We also have neglected the interaction rate of gravitons since they weakly interact with the matter, $\Gamma_g=0$.

\section{CMB photon mixing at the post recombination epoch}

Interactions of photons with medium are in general expressed through the collision integral in order to include the matrix structure of the process. However, for an order of magnitude estimate, the equations of motions for the density operator Eq. \eqref{densityoperator} can be written in a linearized form as follows:
\begin{equation}
\frac{d\hat\rho}{dx}=-i[\hat M,\hat\rho]-\{\hat\Gamma, (\hat\rho-\hat\rho_\textrm{eq})\}
\end{equation}
where the time derivative $d/dt$ of the density operator has been replaced with the spatial derivative $d/dx$ and $\hat\rho_\textrm{eq}$ is the equilibrium density operator of the medium. 


In order to take into account the Universe expansion, we write the spatial derivative of the density operator as $\partial_x=Ha\partial_a$, where $H=\dot a/a$ is the Hubble parameter and $a$ is the cosmological scale factor. Correspondingly the evolution of the density matrix elements is determined by the equations:
\begin{eqnarray}\label{densitysys}
\rho_{\gamma\gamma}' &=&\frac{-2m_{g\gamma}I - \Gamma_\gamma\, (\rho_{\gamma\gamma}-\rho_\textrm{eq})}{Ha},\label{y}\\
\rho_{gg}' &=& \frac{2m_{g\gamma}I}{Ha} ,\label{g}\\
R'&=& \frac{mI-\Gamma_\gamma R/2}{Ha}\label{R1} ,\\
I'&=& \frac{-mR-\Gamma_\gamma I /2 - m_{g\gamma}(\rho_{gg}-\rho_{\gamma\gamma})}{Ha}\label{I},
\end{eqnarray}
where we split the off-diagonal terms of the density matrix as $\rho_{g\gamma}^{*}=\rho_{g\gamma}=R+iI$ 
with $R$ and $I$ being the real part  and the imaginary part, respectively. 
The terms $m_\lambda$ and $m_{g\gamma}$ which enter matrix $M_\lambda$
can be written as functions of the
scale factor as follows (in units of cm$^{-1}$):
\begin{align}
m_{\gamma g}(a) &=  8 \times 10^{-26}\left[\frac{B_i}{1\textrm{G}}\right]\left[\frac{a_i}{a}\right]^2,\\
m_\lambda(a)&=8.75 \times 10^{-34}\left(\frac{B_i}{1 \textrm{G}}\right)^2\left(\frac{f_i}
{\textrm{GHz}}\right)\left(\frac{a_i}{a}\right)^5\label{plas}\\ &-1.7 \times 10^{-8}X_e(a)\left(\frac{\textrm{GHz}}{f_i}\right)\left(\frac{a_i}{a}\right)^2\nonumber
\end{align}
where $B_i$ is the value of the magnetic field at recombination time $t_i$, $a_i$ is the scale factor at recombination time, $f_i=\omega_i/2\pi$ is the initial photon frequency at recombination and $X_e(a)$ is the ionization function of free electrons at the post recombination epoch. On deriving Eq. \eqref{plas} we have expressed the free electron density as $n_e=X_en_B$ where $n_B$ is the total baryon 
number density. The baryon number density can be written as a function of temperature $T$ as $n_B=n_B(t_0)(T/T_0)^3$ where $T_0=2.275$ K is the present day temperature of the CMB photons and the total baryon number density at present is $n_B(t_0)\simeq 2.47\times 10^{-7}$ cm$^{-3}$ \cite{Komatsu:2010fb}. The ionization fraction is not easy to calculate analytically and often 
numerical calculations are used for its determination. 
The evolution of  $X_e$ is determined by the following differential equation \cite{Weinberg:2008}:
\begin{align}\label{eqioniz}
\frac{\mathrm d X_e}{\mathrm d a}&=-\frac{\alpha n_B}{Ha}\left(1+\frac{\beta}{\Gamma_{2s}+8\pi H/\lambda_{\alpha}^3n_B(1-X_e)}\right)^{-1}\\
&\times \left(\frac{SX_e^2+X_e-1}{S}\right),
\end{align} where $\Gamma_{2s}=8.22458$ s$^{-1}$ is the two-photon decay rate of $2s$ hydrogen state, 
$\lambda_{\alpha}=1215.682\times 10^{-8}$ cm is the wavelength of the Lyman $\alpha$ photons, $\alpha(a)$ is the case B recombination 
coefficient,
and $S(a)$ is the coefficient in the Saha equation, $X(1+SX)=1$. Coefficient $\alpha$ depends on the scale factor as \cite{Hummer1994}:
\begin{equation}
\alpha(a)=\frac{1.038\times 10^{-12}a^{0.6166}}{1+0.352a^{-0.53}}\quad \textrm{cm}^3\, \textrm{s}^{-1},
\end{equation}
while  $S(a)$ is equal to
\begin{equation}
S(a)=6.221\times 10^{-19}e^{53.158a}a^{-3/2}.
\end{equation}
Coefficient $\beta$, which is also a function of the scale factor, can be expressed through $\alpha$ as follows
\begin{equation}
\beta(a)=3.9\times 10^{20}a^{-3/2}e^{-13.289a}\alpha\quad\textrm{cm}^{-3}.
\end{equation}
With these parameters Eq. \eqref{eqioniz} is solved numerically and the solution determines the number density of free electrons at the post recombination epoch.

\section{Graviton production at the post recombination epoch: non resonant case}

In the previous section we derived the equations of motions for the elements of the density matrix and in this section we solve them in the case of conversion of the CMB photons into gravitons at the post recombination epoch. To this end we need to know the strength of the magnetic field at the recombination time, $B_i$; the interaction rate of the photons with medium $\Gamma_\gamma$; the ionization function of the plasma as a function of the scale factor $X_e(a)$; and the value of the product $Ha$. 

After recombination, CMB photons weakly interact with matter because the Universe expansion rate $H$ became larger than the interaction rate of the photons $H>\Gamma_\gamma$. However, photons still scattered on matter mainly with the free electrons due to Thomson scattering. The interaction rate in this case is $\Gamma_\gamma=n_e\sigma_T$, where $n_e$ is the free electron number density and $\sigma_T=6.65\times 10^{-25}$ cm$^2$ is the Thomson cross section. In terms of the scale factor the photon interaction rate is given by (in units of cm$^{-1}$)
\begin{equation}
\Gamma_\gamma(a)=2.12\times 10^{-22} X_e(a)\left[\frac{a_i}{a}\right]^3,
\end{equation}
where we used $n_B(t_i)=n_B(t_0)(1+z_i)^3\simeq 320$ cm$^{-3}$ with $1+z_i=1090$.

The value of the magnetic field used in this work is based on upper limits on the angular fluctuations and Faraday rotation of the CMB. Indeed, the presence of large scale magnetic fields can induce CMB anisotropies and measurements of the CMB angular fluctuations \cite{Paoletti:2012bb} constraint the value of the present day magnetic field to be, $B(t_0)\lesssim 3\times 10^{-9}$ G on scales $\lambda_B\sim 1$ Mpc. On the other hand measurement of the Faraday rotation of the CMB polarization \cite{Kahniashvili:2008hx}, limit the present day value of the magnetic field to be $B(t_0)\lesssim 6\times 10^{-8}-2\times 10^{-6}$ G on scales $\lambda_B \sim 0.1-10^3$ Mpc. At the time of recombination the value of the magnetic field is obtained by simply redshifting the present day value of $B(t_0)$, where $B_i=B(t_0)(1+z_i)^2$. 

At the post recombination epoch the Universe was dominated by the nonrelativistic matter and at the late epoch by the vacuum energy. Therefore the value of the product $Ha$ is given by (in what follows we choose the normalization $a_i$=1) 
\begin{equation}
Ha=H(t_i)[\Omega_M/a+\Omega_\lambda a^2]^{1/2},
\end{equation}
where $\Omega_M\simeq 0.3$ is the density parameter of matter; $H^{-1}(t_i)=6.7\times 10^{23}$ cm is the Hubble distance at recombination; and $\Omega_\lambda$ is the density parameter of the vacuum energy.

The maximum probability of graviton production occurs when the oscillation between the photon-graviton system is in resonance. The resonance occurs whenever the function $m_\lambda(a)=0$ which implies 
\begin{equation}\label{resonance}
\omega_\textrm{res}(a)=2.9X_e^{1/2}(a)\left(\frac{1 \textrm{G}}{B_i}\right)a^{3/2}\, \textrm{MeV}.
\end{equation}
As we can see from Eq. \eqref{resonance},  $\omega_\textrm{res}(a)$ depends essentially on the value of the magnetic field at recombination $B_i$ and on the ionization fraction $X_e(a)$. After recombination the ionization function rapidly decreases reaching a constant value of the order $10^{-4}$ and after becomes of the order of unity at the reionization epoch. Since CMB photons are observed at present in the frequency range $0.5-600$ GHz, we can easily conclude that for the present day bounds on $B_i$, CMB photons never cross the resonance energy, Eq. \eqref{resonance}, during their evolution until the present day. Hence, the CMB photons never make resonant oscillations into gravitons.

The system of Eqs. \eqref{densitysys}-\eqref{I} is not easy to solve even numerically because of the fast oscillations of the real and imaginary parts $I, R$ which makes the solutions unstable. However, they can be still solved numerically if one makes some physically reasonable assumptions.  Here we work in the steady state approximation, namely, we assume that $I'=R'=0$ and seek solutions for the real part $R$ and the imaginary part $I$ in terms of $\rho_{\gamma\gamma}$ and $\rho_{gg}$. The steady state approximation is valid as far as the oscillation length of the photon-graviton system $l_\textrm{osc}\propto 1/m_\lambda$ is smaller than the Hubble distance, $l_\textrm{osc}\ll H^{-1}$; see \cite{Dolgov:2002wy} for the case of neutrino oscillations in the cosmological plasma. This condition is very well satisfied at the post recombination, see \cite{Dolgov:2012be}, for the numerical values. 
In this case the system of Eqs.  \eqref{R1} and \eqref{I} becomes
\begin{eqnarray}\label{densitysys1}
R&=& \frac{m_\lambda m_{g\gamma}}{m_\lambda^2+(\Gamma_\gamma/2)^2}(\rho_{\gamma\gamma}-\rho_{gg})\label{R2} ,\\
I&=& \frac{(\Gamma_\gamma/2)m_{g\gamma}}{m_\lambda^2+(\Gamma_\gamma/2)^2}(\rho_{\gamma\gamma}-\rho_{gg})\label{I2}.
\end{eqnarray}

In the nonresonant case ($m_\lambda\neq 0$), the plasma effects dominates over the QED effects and in the energy range of the CMB photons the condition $m_\lambda^2\gg \Gamma_\gamma^2/4$ is very well satisfied. In this case Eqs. \eqref{R2} and \eqref{I2} become
\begin{eqnarray}\label{densitysys1}
R&=& \frac{m_{g\gamma}}{m_\lambda}(\rho_{\gamma\gamma}-\rho_{gg})\label{R3} ,\\
I&=& \frac{(\Gamma_\gamma/2)m_{g\gamma}}{m_\lambda^2}(\rho_{\gamma\gamma}-\rho_{gg})\label{I3}.
\end{eqnarray}
We can clearly see from Eq. \eqref{I3} that the imaginary part is proportional to the interaction rate of the photons as one would have expected from simple arguments. Inserting Eqs. \eqref{R3} and 
\eqref{I3} into Eqs. \eqref{y} and \eqref{g}, we get the following closed system of equations (here we take $\rho_\textrm{eq}=0$)
\begin{eqnarray}\label{sys}
\rho_{\gamma\gamma}' &=&-\frac{\Gamma_\gamma}{Ha}\left[\frac{m_{g\gamma}^2}{m_\lambda^2}(\rho_{\gamma\gamma}-\rho_{gg})+\rho_{\gamma\gamma}\right]\label{y1}\\
\rho_{gg}' &=&  \frac{\Gamma_\gamma}{Ha}\left[\frac{m_{g\gamma}^2}{m_\lambda^2}(\rho_{\gamma\gamma}-\rho_{gg})\right].\label{g1}\nonumber\\
\end{eqnarray}

We solve Eqs. \eqref{y1} and \eqref{g1} numerically and assume the background of CMB photons to be unpolarized at recombination, and therefore we impose the initial conditions $\rho_{\gamma\gamma}(a_i)=2\times 1/2, \rho_{gg}(a_i)=0$ and $I(a_i)=R(a_i)=0$ on the solutions of Eqs. \eqref{y1} and \eqref{g1}. The factor 2 in $\rho_{\gamma\gamma}$ takes into account the two polarization states of the electromagnetic waves (photons).

\begin{figure*}[htbp]
\begin{center}
\includegraphics[scale=1.1]{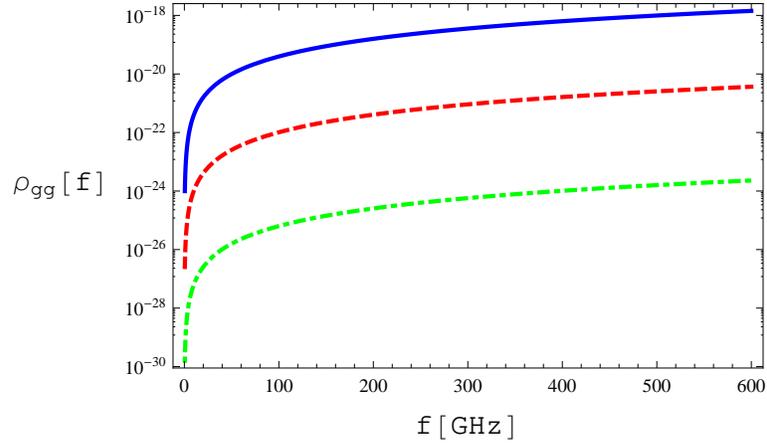}
\caption{Log plot of the graviton production probability $\rho_{gg}$ as a function of the photon/graviton frequency in GHz. In blue (thick) $\rho_{gg}$ is shown for the initial value of the magnetic field at recombination $B_i=2.37$ G; in red (dashed) $\rho_{gg}$ is shown for $B_i=0.12$ G; and in green (dot dashed) $\rho_{gg}$ is shown for $B_i=3\times 10^{-3}$ G.}
\label{Fig1}
\end{center}
\end{figure*}

\begin{figure*}[htbp]
\begin{center}
\includegraphics[scale=1.1]{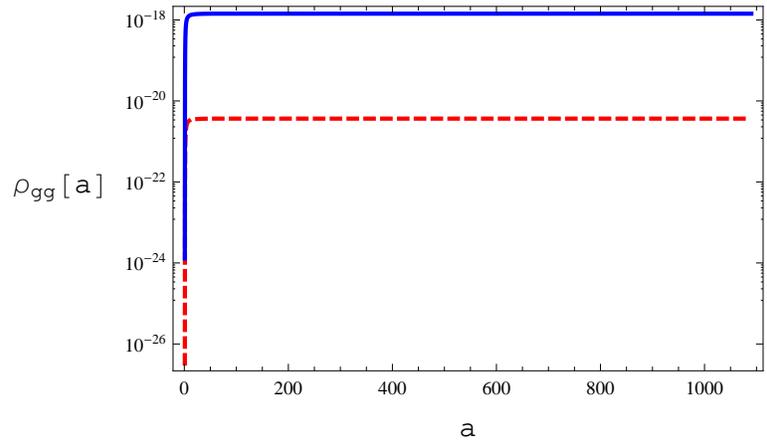}
\caption{Log plot of the graviton production probability $\rho_{gg}$ as a function of the scale factor $a$ for the present day graviton frequency $f=600$ GHz. In blue (thick) $\rho_{gg}$ is shown for the initial value of the magnetic field at recombination $B_i=2.37$ G; and in red (dashed) for the value of the magnetic field $B_i=0.12$ G}
\label{Fig2}
\end{center}
\end{figure*}

In Fig. \ref{Fig1} the graviton production probability, $\rho_{gg}$, as a function of the present day graviton frequency $f$ is shown for various values of the magnetic field at the post recombination epoch. As we can see, the probability increases with frequency depending on the value of the magnetic field at recombination, $B_i$. In our calculations we took into account the fact that the ionization function reached the value of unity at the beginning of the reionization epoch $X_e(a_\textrm{ion}) \sim 1$\cite{Dunkley:2008ie}, where $a_\textrm{ion}\simeq 136$. In Fig. \ref{Fig2} $\rho_{gg}$ as a function of the scale factor $a$ for $f=600$ GHz is shown. The production probability rapidly increases for $a\lesssim 52$ due to a decrease of the density of free electrons and at the onset of the reionization epoch, $a>52$, it reaches an asymptotically constant value of $\rho_{gg}\sim O(10^{-18})$ for $B_i\sim 2$ G and $\rho_{gg}\sim O(10^{-20})$ for $B_i\sim 0.1$ G.

It is interesting at this point to calculate the spectrum of the GWs as a result of the transformation of the CMB photons into gravitons. The energy density of the CMB photons ($\rho_\gamma$) in thermal equilibrium at the temperature $T_0$ in the frequency range $f$ and $f+df$ is given by the Planck distribution law,
\begin{equation}
d\rho_\gamma(f)=\frac{16\pi^2\,f^3}{e^{2\pi f/T_0}-1}\,df.
\end{equation}
The density parameter of CMB photons is $\Omega_\gamma(f)=(1/\rho_c)d\rho_\gamma/d\log f$ and the associated density parameter in GWs is given by
\begin{equation}
\Omega_\textrm{gw}(f)=\Omega_\gamma(f)\,\rho_{gg}(f)=\frac{1}{\rho_c}\frac{16\pi^2f^4}{e^{2\pi f/T_0}-1}\,\rho_{gg}(f),
\end{equation}
where $\rho_c=10.75\, h_0^2$ keV/cm$^{3}$=$4.42\times 10^{50} h_0^2$ Hz$^4$ is the critical energy density.

\begin{figure*}[htbp]

\centering
\mbox{
\subfloat[\label{fig:Fig3}]{\includegraphics[scale=0.9]{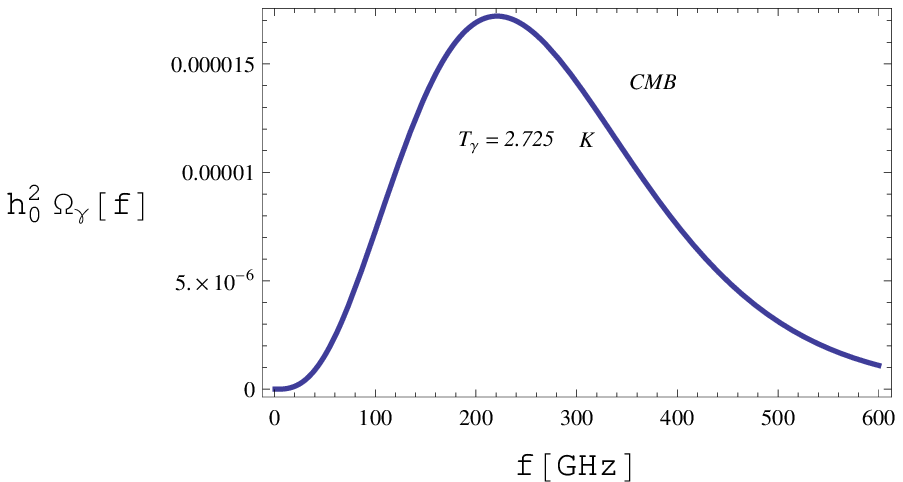}}\qquad

\subfloat[\label{fig:Fig4}]{\includegraphics[scale=0.9]{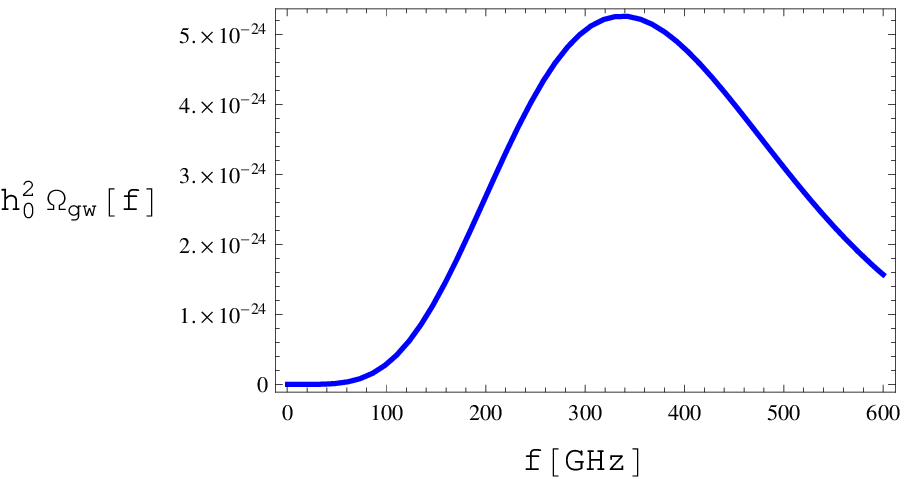}}}

\caption{Plot of the density parameter of CMB photons and the density parameter of the produced gravitons as a function of frequency $f$ in GHz. In (a) the density parameter of CMB for a pure blackbody radiation with temperature $T_\gamma=2.725$ K is shown. In (b) the density parameter of formed gravitons $h_0^2\Omega_\textrm{gw}$ for the value of magnetic field at recombination $B_i=2.37$ G is shown.}
\label{fig:Fig3a}
\end{figure*}

\begin{figure*}[htbp]

\centering

\mbox{

\subfloat[\label{fig:Fig5}]{\includegraphics[scale=0.9]{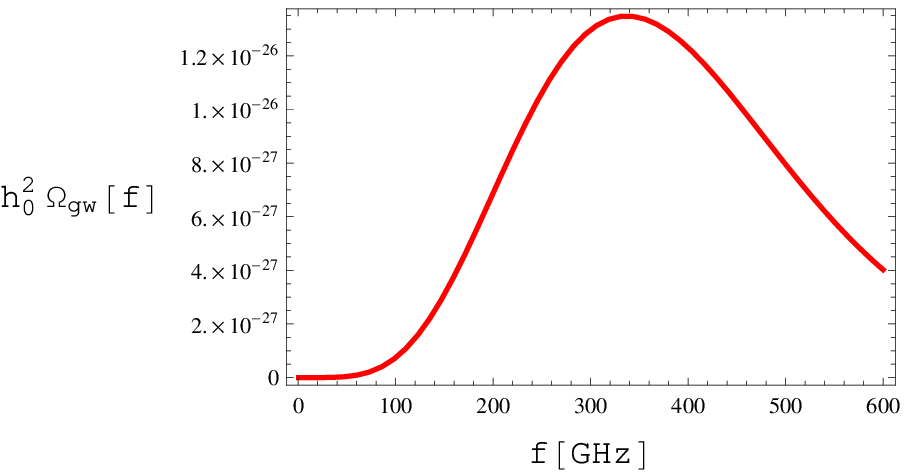}}\qquad
\subfloat[\label{fig:Fig6}]{\includegraphics[scale=0.9]{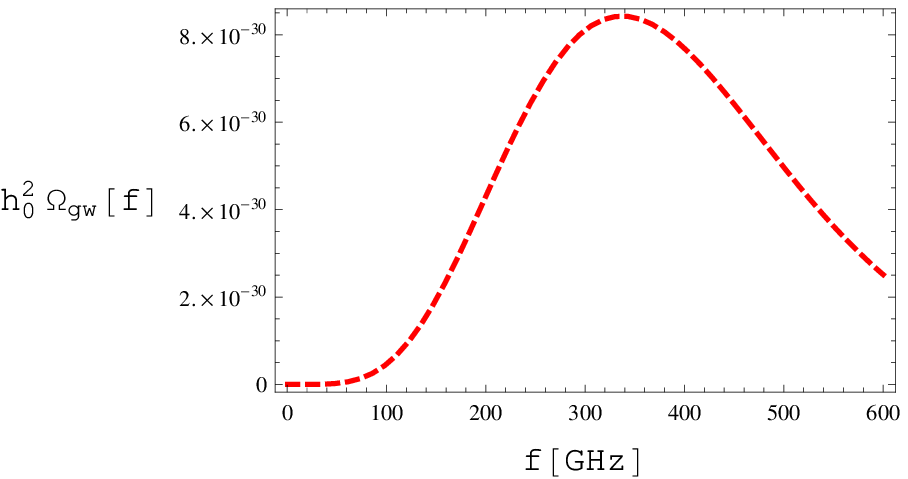}}}
   
\caption{Plot of the graviton density parameter $h_0^2\Omega_\textrm{gw}$ as a function of frequency $f$ in GHz. In (a) the plot for the initial value of the magnetic field at recombination $B_i=0.12$ G is shown and in (b) for $B_i=3$ mG is shown.}
\label{fig:Fig4a}
\end{figure*}

In Fig. \ref{fig:Fig3a} the density parameters of the CMB Fig. \ref{fig:Fig3} and the density parameter of the produced gravitons Fig. \ref{fig:Fig4} for $B_i=2.37$ G are respectively shown. In Fig. \ref{fig:Fig4a} the density parameter in GWs, $h_0^2\Omega_\textrm{gw}$, is shown in the case of the value of the magnetic field at recombination $B_i=0.12$ G Fig. \ref{fig:Fig5} and $B_i=3$ mG Fig. \ref{fig:Fig6}. Apart from the fact that the graviton density parameter is several orders of magnitude smaller than the density parameter of the CMB photons, we can clearly observe from Figs. \ref{fig:Fig3a} and \ref{fig:Fig4a} some differences. First, the graviton spectrum presents a shifted peak frequency with respect to the CMB ($f_\gamma^\textrm{peak}\simeq 220$ GHz) in the high frequency regime with $f_g^\textrm{peak}\simeq 350$ GHz and, second, the blackbody spectrum is not completely preserved.

\section{Discussion and conclusions}

In this work we have discussed the conversion of the CMB photons into gravitons at the post recombination epoch. We calculated the oscillation probability in the density matrix formalism where interaction of the CMB photons with the medium have been taken into account. The conversion probability depends on three main parameters such as the photon frequency $f$, the electronic density of the post recombination plasma $n_e$, and the value of the magnetic field at the post recombination epoch $B$. The graviton production probability $\rho_{gg}$ is independent on the photon polarization states, $+, \times$, thus the produced GWs  background results unpolarized as the CMB and therefore it does not become polarized as a result of the photon to graviton conversion. 

In order to produce the observed CMB temperature anisotropy, $\delta T/T\sim 10^{-5}$, the photon production probability $\rho_{gg}$ must be of the same order $\delta T/T\sim \rho_{gg}$. However, as we can see from Figs. \ref{Fig1} and \ref{Fig2}, $\rho_{gg}$ is smaller than the observed temperature anisotropy by at least 13 orders of magnitude, considering the most favorable case with $B_i\sim 2$ G. This lead us to conclude that the mechanism of the conversion of the CMB photons into gravitons produces a completely negligible temperature anisotropy.

Despite this fact, the discussed mechanism is potentially interesting from the point of view of GW production and their detection. As we have seen, the produced gravitons have the same frequency range as the CMB photons, so this mechanism produces an isotropic background of high frequency of GWs. In Figs. \ref{fig:Fig3a} and \ref{fig:Fig4a} the density parameter in GWs, $h_0^2\Omega_\textrm{gw}$, is shown in the frequency range of $f\sim 0.5-600$ GHz. Its value is in the range of $h_0^2\Omega_\textrm{gw}\sim 10^{-29}-10^{-25}$ depending on the initial value of $B_i$. The spectrum of the produced gravitons is a blackbody like with the peak frequency shifted in the part of the high frequency range by a factor $\sim 1.4$ with respect to the CMB peak frequency. The difference of the graviton spectrum from the blackbody is due to the fact that the graviton production probability depends on the frequency because of the plasma effects. 

The existence of such high frequency background provides another good opportunity to investigate the high frequency range of GWs by the next generation of the high frequency GW detectors. However, since the predicted density parameter in GWs is small and is concentrated in the high frequency part of the spectrum, the detection of such GWs background would be very challenging. Together with the stochastic background of GWs produced during the preheating epoch, see \cite{Buchmuller} for recent calculations, the proposed mechanism falls in the frequency gap between the high frequency part of inflationary models \cite{Maggiore} and the high frequency of GWs emitted by primordial black holes \cite{hi-f-gw}.

\begin{acknowledgements}
The author would like to thank A. D. Dolgov for useful comments and for reviewing the manuscript. 
\end{acknowledgements}

  \end{document}